\documentclass[11pt,a4paper,oneside]{article}
\usepackage[T1]{fontenc}
\usepackage[utf8]{inputenc}
\usepackage[top=3cm, bottom=3cm, left=3cm, right=3cm]{geometry}
\usepackage{amsthm}
\usepackage{authblk}
\usepackage{xspace}
\usepackage{color}
\usepackage{graphicx}
\usepackage[nofiglist,nomarkers]{endfloat}
\usepackage{amsfonts}
\usepackage{subfigure}
\usepackage{mathtools}
\usepackage{newfloat}


\newcommand{\MCMCABC}{\texttt{ABC-MCMC}\xspace}
\newcommand{\ourABC}{\texttt{ABC-PaSS}\xspace}
\newcommand{\WFABC}{\texttt{WF-ABC}\xspace}
\newcommand{\beginsupplement}{%
        \setcounter{table}{0}
        \renewcommand{\thetable}{S\arabic{table}}%
        \setcounter{figure}{0}
        \renewcommand{\thefigure}{S\arabic{figure}}%
     }
\newtheorem{thm}{\bf Theorem}

\def\Sig{\mbox{\boldmath $\Sigma$}}
\def\th{\mbox{\boldmath $\theta$}}
\def\eps{\mbox{\boldmath $\epsilon$}}
\def\mmu{\mbox{\boldmath $\mu$}}
\def\tauu{\mbox{\boldmath $\tau$}}
\def\etaa{\mbox{\boldmath $\eta$}}

\def\bet{\mbox{\boldmath $\beta$}}
\def\xii{\mbox{\boldmath $\xi$}}

\def\s{\mbox{\boldmath $s$}}
\def\c{\mbox{\boldmath $c$}}

\def\C{\mbox{\boldmath $C$}}
\def\D{\mbox{\boldmath $D$}}

\def\P{\mbox{\boldmath $P$}}
\def\T{\mbox{\boldmath $T$}}
\def\I{\mbox{\boldmath $I$}}

\def\I{\mbox{\boldmath $I$}}
\def\d{\mbox{\boldmath $d$}}
\def\t{\mbox{\boldmath $t$}}
\def\0{\mbox{\boldmath $0$}}
\def\F{\mbox{\boldmath $F$}}

\def\P{\mathbb{P}}
\def\E{\mathbb{E}}

\title{Likelihood-free inference in high-dimensional models}
\author[1,2]{Athanasios Kousathanas}
\author[3]{Christoph Leuenberger}
\author[4]{Jonas Helfer}
\author[5]{Mathieu Quinodoz}
\author[5,6]{Matthieu Foll}
\author[1,2]{Daniel Wegmann}
\affil[1]{Department of Biology and Biochemistry, University of Fribourg, Fribourg, Switzerland}
\affil[2]{Swiss Institute of Bioinformatics, Lausanne, Switzerland}
\affil[3]{Department of Mathematics, University of Fribourg, Fribourg, Switzerland}
\affil[4]{Massachusetts Institute of Technology (MIT),Cambridge MA, USA}
\affil[5]{School of Life Sciences, École Polytechnique Fédérale de Lausanne (EPFL), Lausanne, Switzerland}
\affil[6]{International Agency for Research on Cancer, Lyon, France}
\date{}

\begin{document}

\maketitle

\begin{abstract}
Methods that bypass analytical evaluations of the likelihood function have become an indispensable tool for statistical inference in many fields of science. These so-called likelihood-free methods rely on accepting and rejecting simulations based on summary statistics, which limits them to low dimensional models for which the absolute likelihood is large enough to result in manageable acceptance rates. To get around these issues, we introduce a novel, likelihood-free Markov-Chain Monte Carlo (MCMC) method combining two key innovations: updating only one parameter per iteration and accepting or rejecting this update based on subsets of statistics sufficient for this parameter. This increases acceptance rates dramatically, rendering this approach suitable even for models of very high dimensionality. We further derive that for linear models, a one dimensional combination of statistics per parameter is sufficient and can be found empirically with simulations. Finally, we demonstrate that our method readily 
scales to models of very high dimensionality using both toy models as well as by jointly inferring the effective population size, the distribution of fitness effects of new mutations (DFE) and selection coefficients for each locus from data of a recent experiment on the evolution of drug-resistance in Influenza.
\end{abstract}

\renewcommand{\abstractname}{Significance Statement}

\begin{abstract}
The goal of statistical inference is to learn about the parameters of a model that led to the data observed. In complex models, this is often difficult due to a lack of analytical solutions. A popular solution is to replace direct calculations with computer simulations, but the inefficiency of sampling algorithms currently restricts this to low-dimensional models. Here we construct a novel approach that exploits the observation that the information about a parameter is often contained in a subset of the data. This approach readily scales to high dimensions and enables inference in more complex and possibly more realistic models. It allowed us, for instance, to accurately infer parameters underlying the evolution of drug-resistance in Influenza viruses.\end{abstract}

\newpage
The past decade has seen a rise in the application of Bayesian inference algorithms that bypass likelihood calculations with simulations. Indeed, these generally termed likelihood-free or Approximate Bayesian Computation (ABC) \cite{Beaumont2002} methods have a wide range of applications ranging from communications engineering to population genetics \cite{brooks_handbook_2011}. This is because many scientific fields employ complex models for which likelihood calculations are intractable, thus necessitating inference through simulations.

Let us consider a model ${\cal M}$ that depends on $n$ parameters $\th$, creates data $D$ and has the posterior distribution

\begin{eqnarray*}
\pi(\th|D)=\frac{\P(D|\th)\pi(\th)}{\int \P(D|\th)\pi(\th)},
\end{eqnarray*}

where $\pi(\th)$ is the prior and $\P(D|\th)$ is the likelihood function. ABC methods bypass the evaluation of $\P(D|\th)$ by performing simulations with parameter values sampled from $\pi(\th)$ that generate $D$, which in turn is summarized by a set of $m$-dimensional statistics $\s$. The posterior distribution is then evaluated by accepting such simulations that reproduce the statistics calculated from the observed data ($\s_{obs}$)

\begin{eqnarray*}
\pi(\th|\s)=\frac{\P(\s=\s_{obs}|\th)\pi(\th)}{\int \P(\s=\s_{obs}|\th)\pi(\th)}.
\end{eqnarray*}

However for models with $m>>1$ the condition $\s=\s_{obs}$ might be too restrictive and requiring a prohibitively large simulation effort. Therefore, an approximation step can be employed by relaxing the condition $\s=\s_{obs}$ to $\parallel \s-\s_{obs}\parallel \le \delta$, where $\parallel x - y \parallel$ is an arbitrary distance metric between $x$ and $y$ and $\delta$ is a chosen distance (tolerance) below which simulations are accepted. The posterior $\pi(\th|\s)$ is thus approximated by
\begin{eqnarray*}
\pi(\th|\s)=\frac{\P(\parallel\s - \s_{obs}\parallel \le\delta|\th)\pi(\th)}{\int \P(\parallel\s - \s_{obs}\parallel \le\delta|\th)\pi(\th)},
\end{eqnarray*}

An important advance in ABC inference for models of low to moderate dimension was the development of methods coupling ABC with Markov Chain Monte Carlo (MCMC) \cite{marjoram_markov_2003}.These methods allow efficient sampling of the parameter space in regions of high likelihood, thus requiring less simulations to obtain posterior estimates \cite{wegmann_efficient_2009}. The original \MCMCABC  algorithm proposed by Marjoram \textit{et al}. (2003) \cite{marjoram_markov_2003} is:

\begin{enumerate}
\item If now at $\th$ propose to move to $\th'$ according to the transition kernel $q(\th'|\th)$.
\item Simulate $D$ using model ${\cal M}$ with $\th'$ and calculate summary statistics $\s$ for $D$.
\item If $\|\s-\s_{obs}\|\le\delta$ go to step 4 otherwise go to step 1.
\item Calculate the Metropolis-Hastings ratio\\
$$h=h(\th,\th')=\min\left( 1, \frac{\pi(\th')q(\th|\th')}{\pi(\th)q(\th'|\th)}\right).$$
\item Accept $\th'$ with probability $h$ otherwise stay at $\th$. Go to step 1.
\end{enumerate}

The sampling success of \MCMCABC is given by the absolute likelihood values, which are often very low even for relative large tolerance values $\delta$. In such situations, the condition $\parallel \s-\s_{obs}\parallel \le\delta$ will impose a quite rough approximation to the posterior. As a result, the utility of \MCMCABC is limited to models of relatively low dimensionality (typically up to 10 parameters; \cite{blum_approximate_2010,fearnhead_constructing_2012}). The same limitation applies to the more recently developed sequential Monte Carlo sampling methods \cite{Sisson2007a,beaumont_adaptive_2009}.

To this end, three approaches have been suggested to address models of higher dimensionality with ABC. The first approach proposes an expectation propagation approximation to factorize the data space \cite{barthelme_expectation_2014}, which is an efficient solution for situations with high dimensional data, but does not directly address the issue of high dimensional parameter spaces. The second approach formulates the problem using hierarchical models and proposes to first estimate the hyper parameters, and then fixing them when inferring parameters of lower hierarchies individually \cite{bazin_likelihood-free_2010}. The third approach consists of first inferring marginal posterior distributions on low dimensional subsets of the parameter space (either one \cite{nott_approximate_2012} or two dimensional \cite{li_extending_2015}), and then reconstructing the joint posterior distribution from those. The latter two approaches benefit from the lower dimensionality of the statistics space when considering subsets 
of the parameters individually, and hence render the acceptance criterion meaningful. However, they will not recover the true joint distribution if parameters are correlated, which is a common feature of complex models.

\section*{Efficient ABC in high-dimensional models}
Here we introduce a new ABC algorithm that exploits the reduction of dimensionality of the summary statistics when focusing on subsets of parameters, but couples the parameter updates in an MCMC framework. As we prove below, this coupling ensures that our algorithm converges to the true joint posterior distribution even for models of very high dimensions.

Let us define the random variable $\T_i=\T_i(\s)$ as an $m_i$-dimensional function of $\s$. We call $\T_i$ {\it sufficient} for the parameter $\theta_i$ if the conditional distribution of $\s$ given $\T_i$ does not depend on $\theta_i$. More precisely, let $\t_{i,obs}=\T_i(\s_{obs})$. Then
\begin{eqnarray*}\label{def_suff_stat}
\P(\s=\s_{obs}|\T_i=\t_{i,obs},\th) &=& \frac{\P(\s=\s_{obs},\T_i=\t_{i,obs}|\th)}{\P(\T_i=\t_{i,obs}|\th)}\nonumber\\
&=&\frac{\P(\s=\s_{obs}|\th)}{\P(\T_i=\t_{i,obs}|\th)}=: g(\s_{obs},\th_{-i}),
\end{eqnarray*}
where $\th_{-i}=(\theta_1,\ldots,\theta_{i-1},\theta_{i+1},\ldots,\theta_n)$ is $\th$ with the $i$-th component omitted.\\

If sufficient statistics $\T_i$ can be found for each parameter $\theta_i$ and their dimension $m_i$ is substantially smaller than the dimension $m$ of $\s$, then the
\MCMCABC algorithm can be greatly improved with the following algorithm which we denoted ABC with \underline{Pa}rameter \underline{S}pecific \underline{S}tatistics or \ourABC onwards:

\bigskip

The algorithm starts at time $t=1$ and at some initial parameter value $\th^{(1)}$.
\begin{enumerate}
\item Choose an index $i=1,\ldots,n$ according to a probability distribution $(p_1,\ldots,p_n)$ with $\sum p_i = 1$ and all $p_i>0$.
\item At $\th=\th^{(t)}$ propose $\th'$ according to the transition kernel $q_i(\th'|\th)$ where $\th'$ differs from $\th$ only in the $i$-th component:
$$\th'=(\theta_1,\ldots,\theta_{i-1},\theta_i',\theta_{i+1},\ldots,\theta_n).$$
\item Simulate $D$ using model ${\cal M}$ with $\th'$ and calculate summary statistics $\s$ for $D$. Calculate $\t_i=\T_i(\s)$ and $\t_{i,obs}=\T_i(\s_{obs})$.
\item Let $\delta_i$ be the tolerance for parameter $\th_i$. If $\|\t_i-\t_{i,obs}\|\le\delta_i$ go to step 5 otherwise go to step 1.
\item Calculate the Metropolis-Hastings ratio
$$h=h(\th,\th')=\min\left( 1, \frac{\pi(\th')q_i(\th|\th')}{\pi(\th)q_i(\th'|\th)}\right).$$
\item Accept $\th'$ with probability $h$ otherwise stay at $\th$.
\item Increase $t$ by one, save a new parameter value $\th^{(t)}=\th$ and continue at step 1.
\end{enumerate}

Convergence of the MCMC chain is guaranteed by

\begin{thm}\label{thm-algo}
For $i=1..n$, if $\delta_i=0$ and $\T_i$ is sufficient for parameter $\theta_i$ then the stationary distribution of the Markov chain is $\pi(\th|\s=\s_{obs})$.
\end{thm}

The proof for Theorem \ref{thm-algo} is provided in the Appendix.\\

The same types of improvements that have been proposed for \MCMCABC can also be used for \ourABC. For example in order to increase acceptance rate we can relax the assumption $\delta_i=0$ to $\delta_i>0$ and the stationary distribution of the Markov chain is then $\pi(\th|\|\t_i'-\t_i\|_i < \delta_i,i=1,\ldots,n)$ and approximates the posterior distribution $\pi(\th|\s=\s_{obs})$. We can also perform an initial calibration ABC step to find an optimal starting position $\th^{(1)}$, tolerance $\delta_i$ and to adjust the proposal kernel for each parameter \cite{wegmann_efficient_2009}.

\subsection*{Toy model 1: Normal distribution}
We first compared the performance of \ourABC and \MCMCABC under a simple model: the normal distribution with parameters mean ($\mu$) and variance ($\sigma^2$). Given a sample of size $n$, the sample mean ($\bar{x}$) is a sufficient statistic for $\mu$, while both $\bar{x}$ and the sample variance (${S}^2$) are sufficient for $\sigma^2$ \cite{casella_statistical_2002}. For \MCMCABC, we used both $\bar{x}$ and ${S}^2$ as statistics. For \ourABC, we used only $\bar{x}$ when updating $\mu$ and both $\bar{x}$ and ${S}^2$ when updating $\sigma^2$.

\begin{figure}[t!]
\centering
\centerline{\includegraphics[keepaspectratio,scale=0.65]{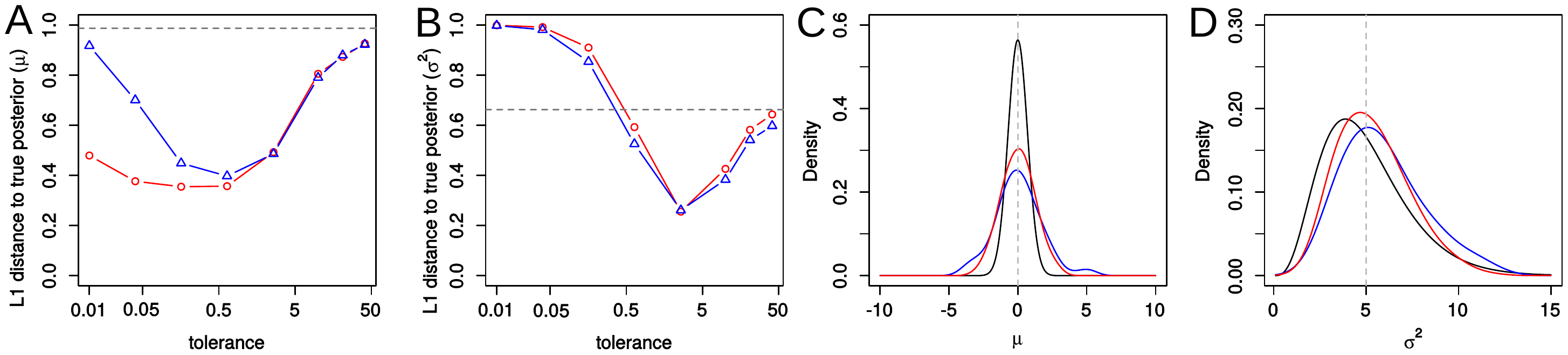}}
\caption{\label{fig_toy1}\textbf{Comparison of performance between \MCMCABC (blue) and \ourABC (red) in estimating the parameters of a normal distribution.} (A,B): the average over 50 chains of the $L_1$ distance between the true and estimated posterior distribution for $\mu$ (A) and $\sigma^2$ (B) for different tolerances. The dashed horizontal line is the $L_1$ distance between the prior and the true posterior distribution. (C,D): The estimated posterior distribution for $\mu$ (C) and $\sigma^2$ (D) using the tolerance that led to the minimum $L_1$ distance from the true posterior (black). The dashed vertical line indicates the true values of the parameters.
}
\end{figure}

We then compared the accuracy between the two algorithms by calculating the total variation distance between the inferred and the true posteriors ($L_1$ distance). We computed $L_1$ under a wide range of tolerances in order to find the tolerance for which each algorithm had the best performance (i.e., minimum $L_1$). As shown in Figure \ref{fig_toy1}, panels A and C, \ourABC produced a more accurate estimation for $\mu$ than \MCMCABC. The two algorithms had similar performance when estimating $\sigma^2$ (Figure \ref{fig_toy1}; B and D).

The normal distribution toy model, although simple, is quite illustrative of the nature of the improvement in performance by using \ourABC over \MCMCABC. Indeed, our results demonstrate that the slight reduction of the summary statistics space by ignoring a single uninformative statistic when updating $\mu$ already results in a noticeable improvement in estimation accuracy. This improvement would not be possible to attain with classic dimension reduction techniques, such as partial least squares (PLS) since the information contained in $\bar{x}$ and ${S}^2$ is irreducible under \MCMCABC.

\section*{Sufficient Statistics in high-dimensional models}
Decreasing the dimensionality of statistics space is  crucial for \ourABC, since we would expect an improvement over \MCMCABC only if we can find per-parameter sufficient statistics of a lower dimension than the total number of parameters. However, choosing summary statistics is not trivial in any ABC application, as too few statistics are insufficient to summarize the data while too many statistics can create an excessively large statistics space that worsens the approximation of the posterior \cite{Beaumont2002,wegmann_efficient_2009, csillery_approximate_2010}. Therefore, several strategies have been developed to reduce the dimensionality of the statistics space \cite{blum_comparative_2013}. For instance, Fearnhead and Prangle \cite{fearnhead_constructing_2012} suggested a method where an initial set of simulations is used to fit a linear model that expresses each parameter $\theta_i$ as a function of $\s$. These functions are then used as statistics in subsequent ABC analysis.

Here we will adopt Fearnhead and Prangle's approach to reduce the dimensionality of statistics space to a single combination of statistics per parameter. This choice is motivated by the finding that for a general linear model (GLM), a single linear function is a sufficient statistic for each associated parameter, as we prove in the following.\\

Suppose that, given the parameters $\th$, the distribution of the statistics vector $\s$ is multivariate normal according to the general linear model (GLM)
\begin{equation*}\label{GLM_base}
\s = \c + \C\th + \eps
\end{equation*}
where $\eps \sim {\cal N}(\0,\Sig_s)$ and for any $m \times n$-matrix $\C$. If the prior distribution of the parameter vector is $\th \sim {\cal N}(\th_0,\Sig_\theta)$ then the posterior distribution of $\th$ given $\s_{obs}$ is
\begin{equation}\label{GLM_posterior}
\th | \s_{obs} \sim \cal{N}(\D\d,\D)
\end{equation}
with $\D=(\C' \Sig_s^{-1} \C + \Sig_\theta^{-1})^{-1}$ and $\d=\C' \Sig_s^{-1} (\s_{obs} - \c) + \Sig_\theta^{-1}\th_0$ (see e.g. \cite{leuenberger_bayesian_2010}). We have the

\medskip

\begin{thm}\label{thm-suff}
Let $\c_i$ be the $i$-th column of $\C$ and $\bet_i = \Sig_s^{-1}\c_i$. Moreover, let
\begin{equation*}\label{tau_i}
\tau_i = \tau_i(\s) = \bet_i' \s.
\end{equation*}
Then $\tau_i$ is sufficient for the parameter $\theta_i$ and the collection of statistics
$$
\tauu=(\tau_1, \ldots, \tau_n)'
$$
yields the same posterior (\ref{GLM_posterior}) as $\s$.

\end{thm}

\medskip

The proof for Theorem \ref{thm-suff} is provided in the Appendix. In practice, the design matrix $\C$ is unknown. We can perform an initial set of simulations from which we can infer that:
$$
\mbox{Cov}(\s, \theta_i)=\mbox{Var}(\theta_i)\c_i.
$$

A reasonable estimator for the sufficient statistic $\tau_i$ is then $\hat{\tau}_i = \hat{\bet}_i' \s$ with
\begin{equation}\label{estimator_tau}
\hat{\bet}_i = \hat{\Sig}_s^{-1} \hat{\Sig}_{s\theta_i},
\end{equation}

where $\hat{\Sig}_s$ and $\hat{\Sig}_{s\theta_i}$ for $i=1,\ldots, n$ are the estimated covariances.\\

\subsection*{Toy model 2: General Linear Model (GLM)}
\begin{figure}[t!]
\centering
\centerline{\includegraphics[keepaspectratio,scale=0.6]{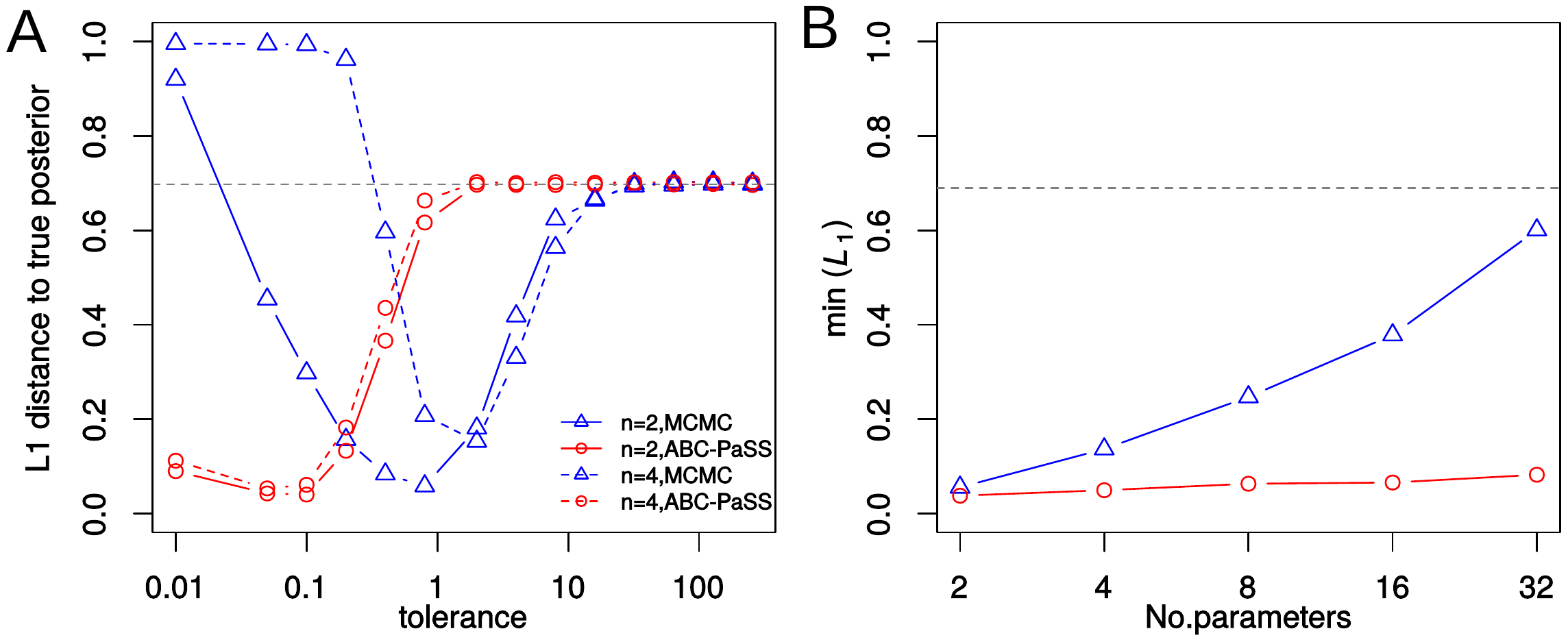}}
\caption{\label{fig_toy2} \textbf{The performance of \MCMCABC (blue) and \ourABC (red) for a GLM model with different numbers of parameters.} (A) the average $L_1$ distance between the true and estimated posterior distribution for different tolerances. Solid and dashed lines are for a GLM with two and four parameters, respectively. (B) the minimum $L_1$ distance from the true posterior over different tolerances for increasing numbers of parameters. (A,B) The dashed line is the $L_1$ distance between the prior and the posterior distribution.
}
\end{figure}

We next compared the performance of \MCMCABC and \ourABC under GLM models of increasing dimensionality $n$. For all models, we constructed the design matrix $\C$ such that all statistics are informative for all parameters, while retaining the total information on the individual parameters regardless of dimensionality (see methods).
For \MCMCABC, we used all statistics $\s\xspace$, while for \ourABC, we employed Theorem \ref{thm-suff} and used a single linear combination of statistics $\tau_i$ per parameter $\theta_i$. As above, we assessed performance of \MCMCABC and \ourABC by calculating the total variation distance ($L_1$) between the inferred and the true posterior distribution. We calculated $L_1$ for several tolerances in order to find the tolerance where $L_1$ was minimal for each algorithm (see Figure \ref{fig_toy2}A for examples with $n=2$ and $n=4$). Since in \MCMCABC distances are calculated in the multi-dimensional statistics space, the optimal tolerance increased with higher dimensionality. This is not the case for \ourABC, because distances are always calculated in one dimension only (Figure \ref{fig_toy2}A).

We found that \MCMCABC performance was good for low $n$, but worsened rapidly with increasing number of parameters, as expected from the corresponding increase in the dimensionality of statistics space (Figure \ref{fig_toy2}B). For a GLM with 32 parameters, approximate posteriors obtained with \MCMCABC differed only little from the prior (Figure \ref{fig_toy2}B). In contrast, performance of \ourABC was unaffected by dimensionality and was better than that of \MCMCABC even in low dimensions (Figure \ref{fig_toy2}B). These results support that by considering low dimensional parameter-specific summary statistics under our framework, ABC inference remains feasible even under models of very high dimensionality, for which current ABC algorithms are not capable of producing meaningful estimates.

\section*{High-dimensional population genetics inference of natural selection and demography}

One of the major research problems in modern population genetics is the inference of natural selection and demographic history, ideally jointly \cite{crisci_recent_2012,bank_thinking_2014}. One way to gain insight into these processes is by investigating how they affect allele frequency trajectories through time in populations, for instance under experimental evolution. Several methods have thus been developed to analyze allele trajectory data in order to infer both locus-specific selection coefficients ($s$) and the effective population size ($N_e$). The modeling framework of these methods assumes Wright-Fisher (WF) population dynamics in a hidden Markov setting to calculate the likelihood of the observed allele trajectories give parameters $N_e$ and $s$ \cite{bollback_estimation_2008,malaspinas_estimating_2012}. In this setting, likelihood calculations are feasible, but very time-consuming, especially when considering many loci at the genome-wide scale \cite{foll_wfabc:_2015}.

To speed-up calculations, Foll \textit{et al.} \cite{foll_wfabc:_2015}
developed an ABC method (\WFABC), adopting the hierarchical ABC framework of Bazin \textit{et al.} \cite{bazin_likelihood-free_2010}. Specifically, \WFABC first estimates $N_e$ based on statistics that are functions of all loci, and then infers $s$ for each locus individually under the inferred value of $N_e$. While \WFABC easily scales to genome-wide data, it suffers from the unrealistic assumption of complete neutrality when inferring $N_e$, which is potentially leading to biases in the inference.

Here we employed \ourABC to infer both $N_e$ and locus-specific selection coefficients jointly. To reduce the dimensionality of the statistics space, we fitted parameter-specific linear combinations as described above, motivated by the frequent and successful application of linear approximations in ABC \cite{Fearnhead2012, wegmann_efficient_2009}. In addition, we first applied a multivariate Box-Cox transformation \cite{box_analysis_1964} to increase linearity between statistics and parameters, as suggested by \cite{wegmann_efficient_2009}, and then assessed the assumption of linearity empirically (Supplementary Figure S1).

\subsubsection*{Perfomance of \ourABC in inferring selection and demography}
\begin{figure}[t!]
\centering
\centerline{\includegraphics[keepaspectratio,scale=0.65]{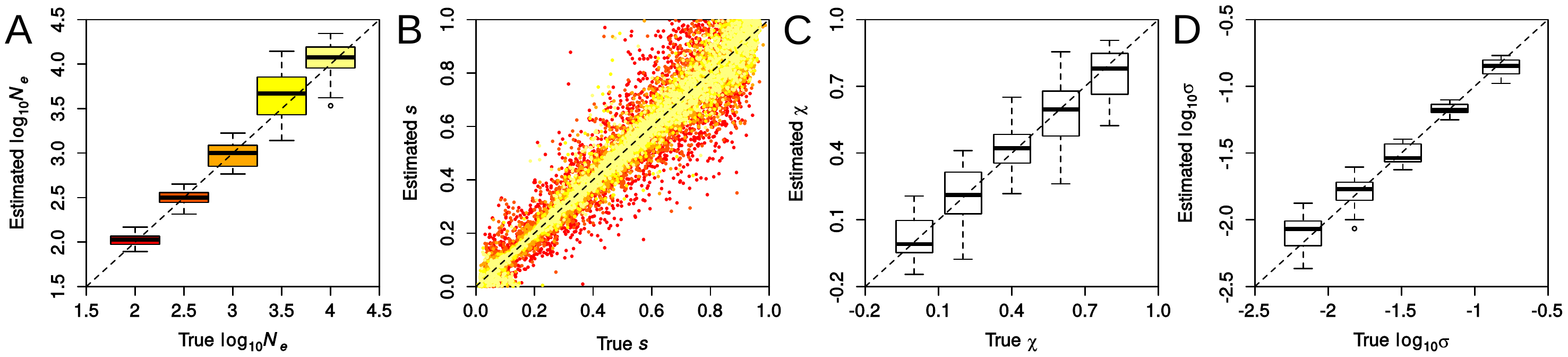}}
\caption{\label{fig_sims}\textbf{Accuracy in inferring demographic and selection parameters by \ourABC.} Shown are the true versus estimated posterior medians for parameters $N_e$ (A), $s$ per locus (B), $\chi$ and $\sigma$ of the Generalized Pareto distribution (C and D, respectively). Boxplots summarize results from 25 replicate simulations, each with 100 loci. Uniform priors over the whole ranges shown were used. (A, B): $N_e$ assumed in the simulations is represented as a color gradient of red (low $N_e$) to yellow (high $N_e$). (C,D): Parameters $\mu$ and $N_e$ were fixed to 0 and $10^{3}$, respectively, $log_{10}\sigma$ was fixed to -1 (C) and $\chi$ was fixed to 0.5 (D).
}
\end{figure}

To examine the performance of \ourABC under the WF model, we inferred $N_e$ and $s$ on sets of 100 loci simulated with varying selection coefficients. We evaluated the accuracy of estimation by comparing the estimated versus the true values of the parameters over 25 replicate simulations. As shown in Figure \ref{fig_sims}A, $N_e$ was estimated well over the whole range of values tested. Estimates for $s$ were on average unbiased and accuracy was, as expected, higher for larger $N_e$ (Figure \ref{fig_sims}B). Note that since the prior on $s$ was $U[0,1]$, these results imply that our approach estimates $N_e$ with high accuracy even when the majority of the simulated loci are under strong selection ($90\%$ of loci had $N_es>10$). Hence, our method allows to relax the assumption of neutrality on most of the loci, which was necessary in previous studies (\cite{foll_wfabc:_2015}).

We next introduced hyper parameters for the distribution of selection coefficients (the so called distribution of fitness effects or DFE). Such hyper parameters are computationally cheap to estimate under our framework, as their updates can be done analytically and do not require simulations. Following \cite{beisel_testing_2007,martin_distribution_2008}, we assumed that the distribution of the locus-specific $s$ is realistically described by a truncated Generalized Pareto Distribution (GPD) with location $\mu=0$ and parameters shape $\sigma$ and scale $\chi$ (Supplementary Figure S2).

We first evaluated the accuracy of estimating $\chi$ and $\sigma$ when fixing the value of the other parameter and found that both parameters are well estimated under these conditions (Figure \ref{fig_sims}, C and D, respectively). Since the truncated GPD of multiple combinations of $\chi$ and $\sigma$ is very similar, these parameters are not always identifiable. This renders the accurate joint estimation of both parameters difficult (Supplementary Figure 3B-C). However, despite the reduced accuracy on the individual parameters, we found the overall shape of the GPD to be well recovered (Supplementary Figure 3D-F). Also, $N_e$ was estimated with high accuracy for all combinations of $\chi$ and $\sigma$ (Supplementary Figure 3A).

\subsubsection*{Analysis of Infuenza data}
We applied our approach to data from a previous study \cite{foll_influenza_2014} where cultured canine kidney cells infected with the Influenza virus were subjected to serial transfers for several generations. In one experiment, the cells were treated with the drug Oseltamivir, and in a control experiment they were not treated with the drug. To obtain allele frequency trajectories of all sites of the Infuenza virus genome (13.5 Kbp), samples were taken and sequenced every 13 generations with pooled population sequencing. The aim of our application was to identify which viral mutations rose in frequency during the experiment due to natural selection and which due to drift and to investigate the shape of the DFE for the control and drug-treated viral populations.\\

\begin{figure}[t!]
\centering
\centerline{\includegraphics[keepaspectratio,scale=0.65]{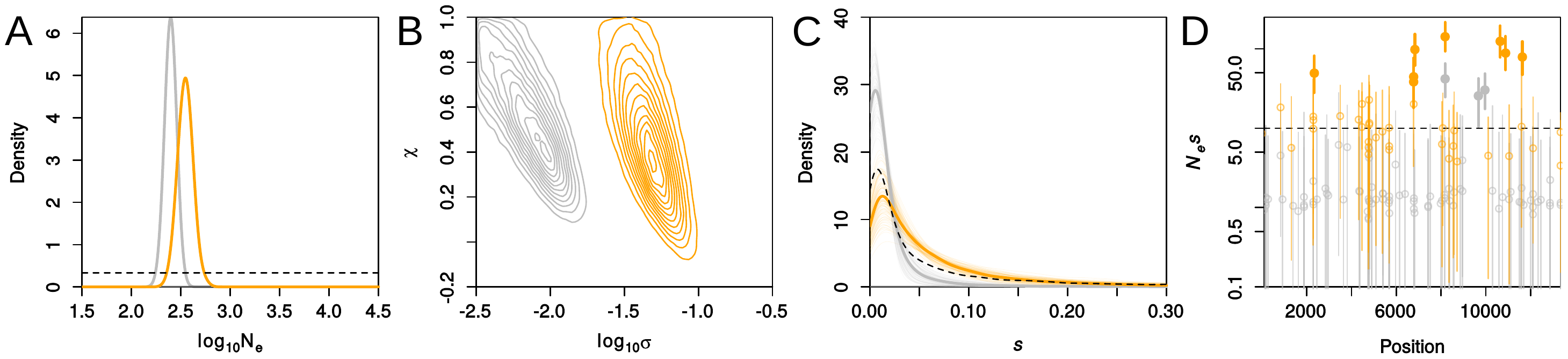}}
\caption{\label{fig_influenza}\textbf{ Inferred demography and selection for experimental evolution of Infuenza.} We show results for the no-drug (control) and drug-treated Influenza in grey and orange, respectively. Shown are the posterior distributions for $log_{10}N_e$ (A) and $log_{10}\sigma$ and $\chi$ (B). In panel C, we plotted the modal distribution of fitness effects (DFE) with thick lines by integrating over the posterior of its parameters. The thin lines represent the DFEs obtained by drawing 100 samples from the posterior of $\sigma$ and $\chi$. Dashed lines in panels A and C correspond to the prior distributions. In panel D, the posterior estimates for $Nes$ per locus versus the position of the loci in the genome are shown. Open circles indicate non-significant loci whereas closed, thick circles indicate significant loci ($P(N_es>10)>0.95$, dashed line).}

\end{figure}

Following \cite{foll_influenza_2014}, we filtered the raw data to contain loci for which sufficient data was available to calculate the summary statiatics considered here (see Methods). There were 86 and 42 such loci for the control and drug-treated experiment, respectively (Supplementary Figure S4).

We then employed \ourABC to estimate $N_e$, $s$ per site and the parameters of the DFE. We obtained a low estimate for $N_e$ (posterior medians 350 for drug-treated and 250 for control Influenza; Figure \ref{fig_influenza}A), which is expected given the bottleneck that the cells were subjected to in each transfer. While we obtained similar estimates for the $\chi$ parameters for the drug-treated and for the control Influenza (posterior medians 0.44 and 0.56, respectively), the $\sigma$ parameter was estimated to be much higher for the drug-treated than for the control Influenza (posterior medians 0.047 and 0.0071, respectively, Figure \ref{fig_influenza}B). The resulting DFE was thus very different for the two conditions: the DFE for the drug-treated Influenza had a much heavier tail than the control (Figure \ref{fig_influenza}C). Posterior estimates for $N_es$ per mutation also indicated that the drug-treated Influenza had more mutations under strong positive selection than the control (19\% versus 3.5\% of 
loci had $P(N_es>10)>0.95$, respectively; Figure \ref{fig_influenza}D, Supplementary Figure S4). These results indicate that the drug treatment placed the Influenza population away from a fitness optimum, thus increasing the number of positively selected mutations with large effect sizes. Presumably these mutations confer resistance to the drug thus helping Influenza to reach a new fitness optimum.

Our results for Influenza were qualitatively similar to those obtained by \cite{foll_influenza_2014}. We obtained slightly larger estimates for $N_e$ (350 versus 226 for drug-treated and 250 versus 176 for control Influenza). Our estimates for the parameters of the GPD were substantially different than \cite{foll_influenza_2014} but resulted in qualitatively similar overall shapes of the DFE for both drug-treated and control experiments. These results underline the applicability of our method to a high-dimensional problem. In contrast to \cite{foll_influenza_2014} who performed estimations in a 3-step approach, combining a moment-based estimator for $N_e$, ABC for $s$ and a maximum likelihood approach for the GPD, our Bayesian framework allowed us to perform joint estimation and to obtain posterior distributions for all parameters in a single step.

\section*{Conclusion}
Due to the difficulty to find analytically tractable likelihood solutions, statistical inference was often limited to models that made substantial approximations of reality. To address this problem, so-called likelihood-free approaches have been introduced that bypass the analytical evaluation of the likelihood function with computer simulations. While full-likelihood solutions generally have more power, likelihood-free methods have been used in many fields of science to overcome undesired model assumptions.

Here we developed and implemented a novel likelihood-free, Markov chain Monte Carlo (MCMC) inference framework that scales naturally to high dimensions. This framework takes advantage of the observation that the information about one model parameter is often contained in a subset of the data, by integrating two key innovations: first, only a single parameter is updated at the time, and that update is accepted based on a subset of summary statistics sufficient for this parameter. We proved that this MCMC variant converges to the true joint posterior distribution under the standard assumptions. We further derived that for linear models, a one-dimensional function of summary statistics per parameter is sufficient. 

We then demonstrated the power of our framework through the application to multiple problems. First, our framework led to more accurate inference of the mean and standard deviation of a normal distribution than standard likelihood-free MCMC, suggesting that our framework is already competitive in models of low dimensionality. In high dimensions, the benefit was even more apparent. When applied to the problem of inferring parameters of a general linear model (GLM), for instance, we found our framework to be insensitive to the dimensionality, resulting in a performance similar to analytical solutions both in low and high dimensions. Finally, we used our framework to address the difficult and high-dimensional problem of inferring demography and selection jointly from genetic data. Specifically, and through simulations and an application to experimental data, we show that our framework enables the accurate joint estimation of the effective population size, the distribution of fitness effects of segregating 
mutations, as well as locus-specific selection coefficients from time series data. 


\newpage
\section*{Materials and methods}

\subsection*{Implementation}
We implemented the proposed \ourABC framework into a new version of the software package \texttt{ABCtoolbox} \cite{Wegmann2010a}, which will be made available at the authors website and will be described elsewhere.

\subsection*{Toy models: Normal distribution}

We performed simulations to assess the performance of \MCMCABC and \ourABC in estimating $\theta_1=\mu$ and $\theta_2=\sigma^2$ for a univariate normal distribution. We used the sample mean $\bar{x}$ and sample variance $S^2$ of samples of size $n$ as statistics. Recall that for non-informative priors the posterior distribution for $\mu$ is ${\cal N}(\bar{x}, S^2/n)$ and the posterior distribution for $\sigma^2$ is $\chi^2$-distributed with $n-1$ degrees of freedom. As $\mu$ and $\sigma^2$ are independent, we get the posterior density
$$
\pi(\mu, \sigma^2) = \phi_{\bar{x}, S^2/n}(\mu) \cdot \frac{n-1}{S^2} f_{\chi^2; n-1}\left(\frac{n-1}{S^2} \sigma^2\right).
$$
In our simulations the sample size was $n=10$ and the true parameters were given by $\mu=0$ and $\sigma^2=5$. We performed 50 MCMC chains per simulation and chose effectively non-informative  priors for $\mu \sim U[-10,10]$ and $\sigma^2 \sim U[0.1,15]$. Our simulations were performed for a wide range of tolerances (from 0.1 to 0.9) and proposal ranges (from 0.1 to 0.9). We did this exhaustive search in order to identify the combination of these tuning parameters that allows \MCMCABC and \ourABC to perform best in estimating $\mu$ and $\sigma^2$. We then recorded the minimum total variation distance ($L_1$) between the true and estimated posteriors over these sets of tolerances and ranges and compared it between \MCMCABC and \ourABC.

\subsection*{Toy models: GLM}
We considered linear models with $m$ statistics $\s$ being a linear function of $n=m$ parameters $\th$:
\begin{equation*}\label{GLM}
\s = \C\th + \eps,\quad \eps \sim {\cal N}({ \0}, \I),
\end{equation*}
where $\C$ is a square design matrix and the vector of errors $\eps$ is multivariate normal. Under non-informative priors for the parameters $\th$, their posterior distribution is multivariate normal
$$
\th | \s \sim {\cal N}\left((\C' \C)^{-1}\C'\s,(\C' \C)^{-1}\right).
$$

We set up the design matrices $\C$ in a cyclic manner to allow all statistics to have information on all parameters but their contributions to differ for each parameter, namely we set $\C = {\bf B} \cdot \det({\bf B}'{\bf B})^{-1/2n}$ where
$$
{\bf B}=\left(
\begin{array}{ccccc}
1/n & 2/n & 3/n & \ldots & n/n\\
n/n & 1/n & 2/n  & \ldots & n-1/n\\
\vdots & \vdots & \vdots & \ddots & 2/n\\
2/10 & 3/10 & 4/10 & \ldots & 1/10
\end{array}\right).
$$
The normalization factor in the definition of $\C$ was chosen such that the determinant of the posterior variance is constant and thus the widths of the marginal posteriors are comparable independently of the dimensionality $n$. We used all statistics for \MCMCABC and calculated a single linear combination of statistics per parameter for \ourABC according to Theorem \ref{thm-suff}. For the estimation, we assumed that $\th = \mathbf{0}$ and the priors are uniform $U[-100,100]$ for all parameters, which are effectively non-informative. We started the MCMC chains at a normal deviate $N(\th,0.01 \I)$, i.e. around the true values of $\th$.  To ensure fair comparisons between methods, we performed simulations of 50 chains for a variety of tolerances (from 0.1 to 0.9) and proposal ranges (from 0.1 to 0.9) to choose the combination of these tuning parameters at which each method performed best. We run all our MCMC chains for $10^5$ iterations per model parameter to account for model complexity.

\subsection*{\ourABC for estimating selection and demography}
\subsubsection*{Model}
Consider a vector  $\xii$ of observed allele trajectories (sample allele frequencies) over $l=1, \ldots, L$ loci, as is commonly obtained in studies of experimental evolution. We assume these trajectories to be the result of both random drift as well as selection, parameterized by the effective population size $N_e$ and locus-specific selection coefficients $s_l$, respectively, under the classic Wright-Fisher model with allelic fitnesses $1$ and $1+s_l$. We further assume the locus-specific selection coefficients $s_l$ follow a distribution of fitness effects (DFE) parameterized as a Generalized Pareto distribution (GPD) with mean $\mu=0$, shape $\chi$ and scale $\sigma$. Our goal is thus to estimate the joint posterior distribution
\begin{equation*}
\pi(N_e, s_1, \ldots, s_L, \chi, \sigma|\xii)\propto \prod_{l=1}^{L} {\big[\P(\xi_l|N_e,s_l)\pi(s_l|\chi, \sigma)}\big]\pi(N_e)\pi(\chi)\pi(\sigma)
\end{equation*}

To apply our \ourABC framework to this problem, we approximate the likelihood term $\P(\xi_l|N_e,s_l)$ numerically with simulations, while updating the hyper-parameters $\chi$ and $\sigma$ analytically.

\subsubsection*{Summary statistics}
To summarize the data $\xii$, we used statistics originally proposed by Foll \textit{et al.} \cite{foll_wfabc:_2015}. Specifically, we first calculated for each locus individually a measure of the difference in allele frequency between consecutive time points as:
\begin{equation*}
Fs'=\frac{1}{t}\frac{Fs[1-1/(2\widetilde{n})]-2/\widetilde{n}}{(1+Fs/4)[1-1/(n_y)]},
\end{equation*}
where
\begin{equation*}
 Fs=\frac{(x-y)^2}{z(1-z)},
 \end{equation*}
 $x$ and $y$ are the minor allele frequencies separated by $t$ generations, $z=(x+y)/2$ and $\widetilde{n}$ is the harmonic mean of the sample sizes $n_x $ and $n_y$ . We then summed the $Fs'$ values of all pairs of consecutive time points with increasing and decreasing allele frequencies into $Fs'i$ and $Fs'd$, respectively \cite{foll_wfabc:_2015}. Finally, we followed \cite{aeschbacher_novel_2012} and calculated boosted variants of the two statistics in order to take more complex relationships between parameters and statistics into account. The full set of statistics used per locus were $\F_l$ = \{$Fs'i_l$, $Fs'd_l$, $Fs'i_l^2$, $Fs'd_l^2$, $Fs'i_l\times Fs'd_l$\} .

 We next calculated parameter-specific linear combinations for $N_e$ and locus-specific $s_l$ following the procedure developed above. To do so, we simulated allele trajectories of a single locus for different values of $N_e$ and $s$ sampled from their prior. We then calculated $\F_l$ for each simulation and performed a boxcox transformation to linearize the relationships between statistics and parameters \cite{box_analysis_1964,wegmann_efficient_2009}. We then fit a linear model as outlined in Equation \ref{estimator_tau} to estimate the coefficients of an approximately sufficient linear combination of $\F$ for each parameter $N_e$ and $s$. This resulted in $\tau_{s}(\F_l)=\bet_s\F_l$ and $\tau_{N_e}(\F_l)=\bet_{N_e}\F_l$. To combine information across loci when updating $N_e$, we then calculated
 \begin{equation*}
\tau_{N_e}(\F)=\sum_{l=1}^{L} {\bet_{N_e}\F_l},
 \end{equation*}
 where $\F=\{\F_1, \ldots, \F_L\}$.  In summary, we used the ABC approximation
\begin{equation*}
\P(\xi_j|N_e, s_j) \approx \P(\|\tau_s(\F_l)-\tau_s(\F_{l_{obs}})\| < \delta_{s_l}, \|\tau_{N_e}(\F)-\tau_{N_e}(\F_{obs})\|< \delta_{N_e})|N_e,s_j).
\end{equation*}

\subsection*{Simulations and Application}
We applied our framework to allele frequency data for the whole Influenza H1N1 genome obtained in a recently published evolutionary experiment \cite{foll_influenza_2014}. In this experiment, Influenza A/Brisbane/59/2007 (H1N1) was serially amplified on Madin-Darby canine kidney (MDCK) cells for 12 passages of 72 hours each, corresponding to roughly 13 generations (doublings). After the three initial passages, samples were passed either in the absence of drug, or in the presence of increasing concentrations of the antiviral drug oseltamivir. At the end of each passage, samples were collected for whole genome high throughput population sequencing. We obtained the raw data from http://bib.umassmed.edu/influenza/ and, following the original study \cite{foll_influenza_2014}, we downsampled it to 1000 haplotypes per timepoint and filtered it to contain only loci for which sufficient data was available to calculate the $Fs'$ statistics. Specifically, we included all loci
with an allele frequency $\ge2\%$ at $\ge2$ timepoints. There were 86 and 42 such loci for the control and drug-treated experiment, respectively. Further, we restricted our analysis of the data of the drug-treated experiment to the last nine time points during which drug was administered.

We performed all our Wright-Fisher simulations with in-house C++ code implemented as a module of \texttt{ABCtoolbox}. We simulated 13 generations between timepoints and a sample of size 1000 per timepoint. We set the prior for $N_e$ uniform on the $\log_{10}$ scale such that $\log_{10}(N_e) \sim U[1.5,4.5]$ and for the parameters of the GPD $\chi \sim U[-0.2,1]$ and for $\log_{10}(\sigma) \sim U[-2.5,-0.5]$. For the simulations where no DFE was assumed, we set the prior of $s \sim U[0,1]$.

As above, we run all our \ourABC chains for $10^5$ iterations per model parameter to account for model complexity. To ensure fast convergence, the \ourABC implementation benefited from an initial calibration step we originally developed for \MCMCABC and implemented in \texttt{ABCtoolbox} \cite{wegmann_efficient_2009}. Specifically, we first generated 10,000 simulations with values drawn randomly from the prior. For each parameter, we then selected the 1\% subset of these simulations with the smallest distances to the observed data based on the linear-combination specific for that parameter. These accepted simulations were used to calibrate three important metrics prior to the MCMC run: first, we set the parameter-specific tolerances $\delta_i$ to the largest distance among the accepted simulations. Second, we set the width of the parameter-specific proposal kernel to half of the standard deviation of the accepted parameter values. Third, we chose the starting value of the chain for each parameter as the
accepted simulation with smallest distance. Each chain was then run for 1,000 iterations, and new starting values were chosen randomly among the accepted calibration simulations for those parameters for which no update was accepted. This was repeated until all parameters were updated at least once.

\section*{Appendix}
{\bf Proof for Theorem 1.} The transition kernel ${\cal K}(\th,\th')$ associated with the Markov chain is zero if $\th$ and $\th'$ differ in more than one component. If $\th_{-i}=\th_{-i}'$ for some index $i$, then we have
\begin{equation}\label{transition_kernel}
{\cal K}(\th,\th')=p_i \rho_i(\th,\th')+(1-r(\th))\delta_{\th}(\th')
\end{equation}
where $\rho_i(\th,\th')=q_i(\th'|\th)\P(\T=\t_{i,obs}|\th')h(\th,\th')$, $\delta_{\th}$ is the Dirac mass in $\th$, and
$$
r(\th)=\sum\limits_{i=1}^n p_i \int \rho_i(\th,\th')d\theta_i.
$$
We may assume without loss of generality that
$$
\frac{\pi(\th')q_i(\th|\th')}{\pi(\th)q_i(\th'|\th)} \leq 1.
$$
From (\ref{def_suff_stat}) we conclude
$$
\P(\s=\s_{obs}|\th)=\P(\T_i=\t_{i,obs}|\th) g_i(\s_{obs},\th_{-i}).
$$
Setting
$$
c:= \left( \int \P(\s=\s_{obs}|\th)\pi(\th)d\th \right)^{-1}
$$
and keeping in mind that $\th_{-i}=\th_{-i}'$ and $h(\th',\th)=1$, we get
\begin{eqnarray*}
\pi(\th|\s=\s_{obs})\rho_i(\th,\th') &=& \pi(\th|\s=\s_{obs})q(\th'|\th)\P(\T_i=\t_{i,obs}|\th')h(\th,\th')\\
&=& c \ \P(\s=\s_{obs}|\th) \pi(\th)q_i(\th'|\th)\P(\T_i=\t_{i,obs}|\th')\frac{\pi(\th')q_i(\th|\th')}{\pi(\th)q_i(\th'|\th)}\\
&=& c \ \P(\T_i=\t_{i,obs}|\th)g_i(\s_{obs},\th_{-i})\P(\T_i=\t_{i,obs}|\th')\pi(\th')q_i(\th|\th')\\
&= &c \ \P(\T_i=\t_{i,obs}|\th')g_i(\s_{obs},\th_{-i}')\P(\T_i=\t_{i,obs}|\th)\pi(\th')q_i(\th|\th')\\
&=& c \ \P(\s=\s_{obs}|\th')\pi(\th')\P(\T_i=\t_{i,obs}|\th)\pi(\th')q_i(\th|\th')h(\th',\th)\\
&=& \pi(\th'|\s=\s_{obs})\rho_i(\th',\th).
\end{eqnarray*}
From this and equation (\ref{transition_kernel}) follows readily that the transition kernel ${\cal K}(\cdot,\cdot)$ satisfies the detailed balanced equation
$$
\pi(\th|\s=\s_{obs}){\cal K}(\th,\th')=\pi(\th'|\s=\s_{obs}){\cal K}(\th',\th)
$$
of the Metropolis-Hastings chain. $\hfill\Box$

\bigskip

{\bf Proof for Theorem 2.} It is easy to check that the mean of $\tau_i$ is $\mu_i = \t_i' (\C\th+\c)$ and its variance is $\sigma_i^2 = \t_i' \Sig_s \t_i$. The covariance between $\s$ and $\tau$ is given by
\begin{eqnarray*}
\Sig_{s\tau} &= & \E\left((\s - \C\th - \c)(\tau_i - \mu_i)\right)\\
&=& \E\left(\eps \eps' \t_i \right) = \Sig_s \t_i.
\end{eqnarray*}
Consider the conditional multinormal distribution $\s | \tau_i$. Using the well-known formula for the variance and the mean of a conditional multivariate normal (see e.g. \cite{bilodeau_theory_2008}, p. 63), we get that the  covariance of $\s | \tau_i$ is given by
$$
\Sig_{s|\tau} = \Sig_s - \sigma_i^{-2} \Sig_{s\tau}\Sig_{s\tau}'
$$
and thus is independent of $\th$. The mean of $\s | \tau_i$ is
$$
\mmu_{s|\tau} = \C\th + \c + \sigma_i^{-2} \Sig_{s\tau} \t_i' \left(\s - \C\th - \c \right).
$$
The part of this expression depending on $\theta_i$ is
$$
\left( \I - \frac{\Sig_s \t_i \t_i'}{\t_i' \Sig_s \t_i}\right) \c_i \theta_i.
$$
Inserting $\t_i = \Sig_s^{-1}\c_i$ we obtain
$$
\left(\ c_i - \frac{\Sig_s \Sig_s^{-1}\c_i \c_i' \Sig_s^{-1} \c_i}{\c_i' \Sig_s^{-1}\Sig_s \Sig_s^{-1} \c_i} \right) \theta_i = (\c_i - \c_i)\theta_i = \0.
$$
Thus the distribution of $\s | \tau_i$ is independent of $\theta_i$ and hence $\tau_i$ is sufficient for $\theta_i$.

\medskip

To prove the second part of the theorem, we observe that $\tauu$ is given by the linear model
$$
\tauu=\C'\Sig_s^{-1} \s = \C'\Sig_s^{-1} \C \th + \C'\Sig_s^{-1}\c + \etaa
$$
with $\etaa = \C'\Sig_s^{-1} \eps$. Using $\mbox{Cov}(\etaa)= \C'\Sig_s^{-1}\C$ we get for the posterior variance
$$
\left(\C'\Sig_s^{-1}(\C'\Sig_s^{-1}\C)^{-1}\C'\Sig_s^{-1}\C + \Sig_\theta^{-1}\right)^{-1}=(\C' \Sig_s^{-1} \C + \Sig_\theta^{-1})^{-1}=\D.
$$
Similarly one sees that the posterior mean is $\D\d$. $\hfill\Box$

\bigskip

\bibliographystyle{pnas}
\section*{Acknowledgments}
We thank Pablo Duchen and the groups of Laurent Excoffier and Jeffrey Jensen for comments and discussion on this work. This study was supported by Swiss National Foundation grant no 31003A\_149920 to D.W.
\newpage
\bibliography{Suffstat_draft_final_arxiv}
\newpage
\processdelayedfloats

\beginsupplement
\section*{Supplementary Material}

\begin{figure}[H]
\centering
\centerline{\includegraphics[keepaspectratio,scale=0.65]{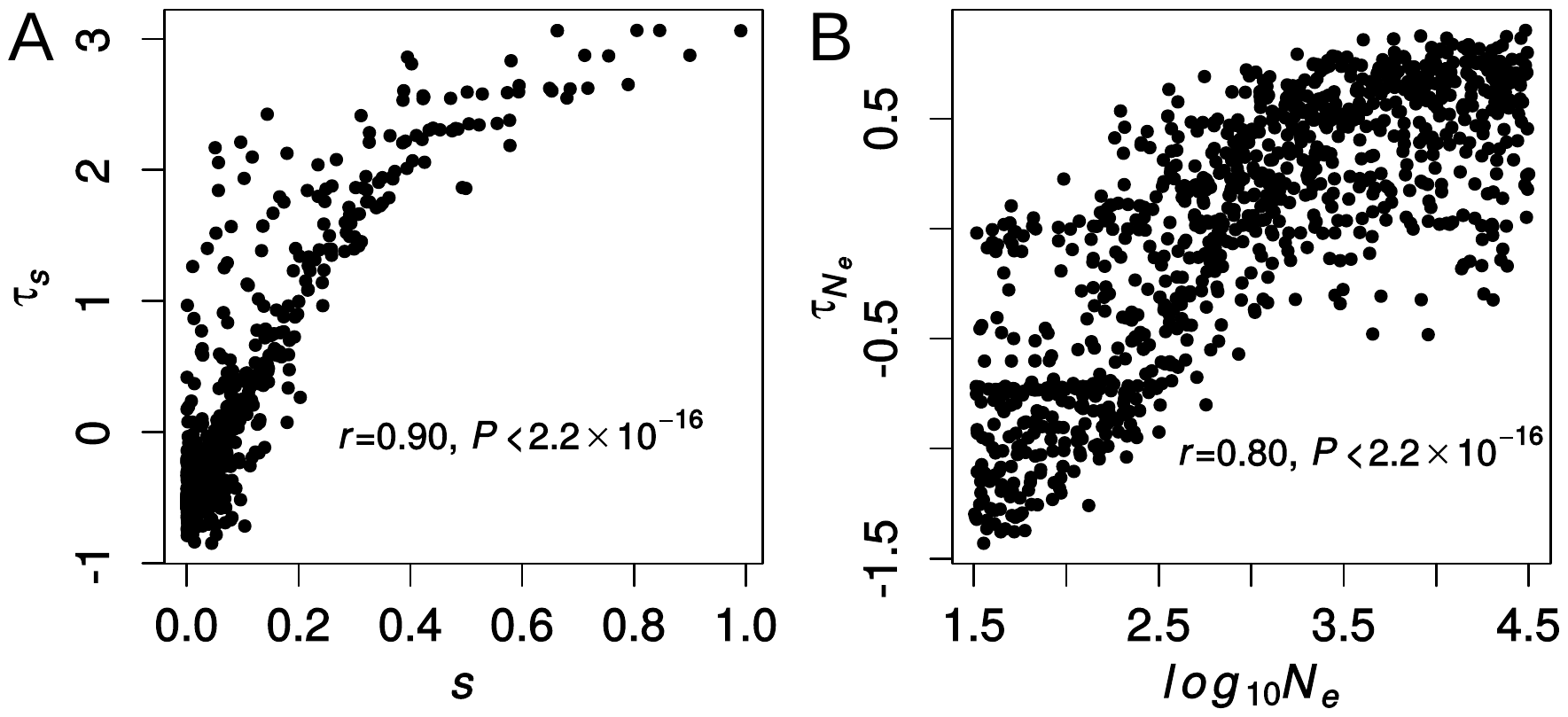}}
\caption{\label{fig_linearcombs}\textbf{Relationship between parameters and linear combinations.} The relationship between the parameters $s$ and $log_{10}N_e$ with respective linear combinations of statistics $\tau_{s}$ and $\tau_{N_e}$ calculated for a set of $10^4$ simulations and assuming priors used for the Influenza application presented in the main text. The Pearson correlation coefficient ($r$) and the corresponding $P$-value are shown in each panel.}
\end{figure}

\newpage

\begin{figure}[!htb]
\centering
\centerline{\includegraphics[keepaspectratio,scale=0.5]{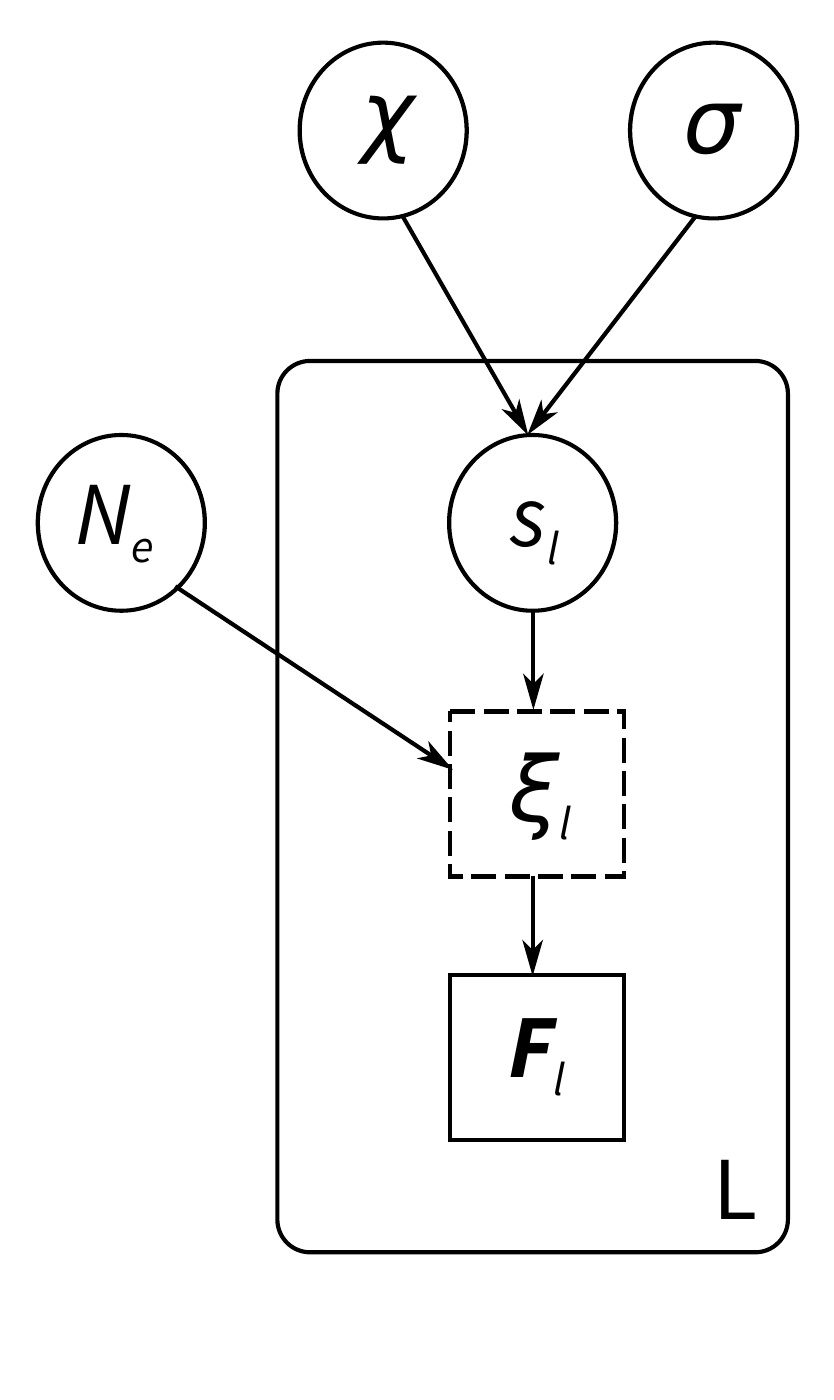}}
\caption{\label{fig_DAG}\textbf{Directed acyclic graph describing the Wright-Fisher model examined in this study.} Solid circles represent parameters to be estimated. The dashed square represents the full data, which is summerized here by a vector of statistics $\boldsymbol{F}_l$, indicated by a solid square. Nodes contained in the plate are repeated  for each locus $l \in \{1, \ldots, L\}$ times.}
\end{figure}
\newpage

\begin{figure}[!htb]
\centering
\centerline{\includegraphics[keepaspectratio,scale=0.5]{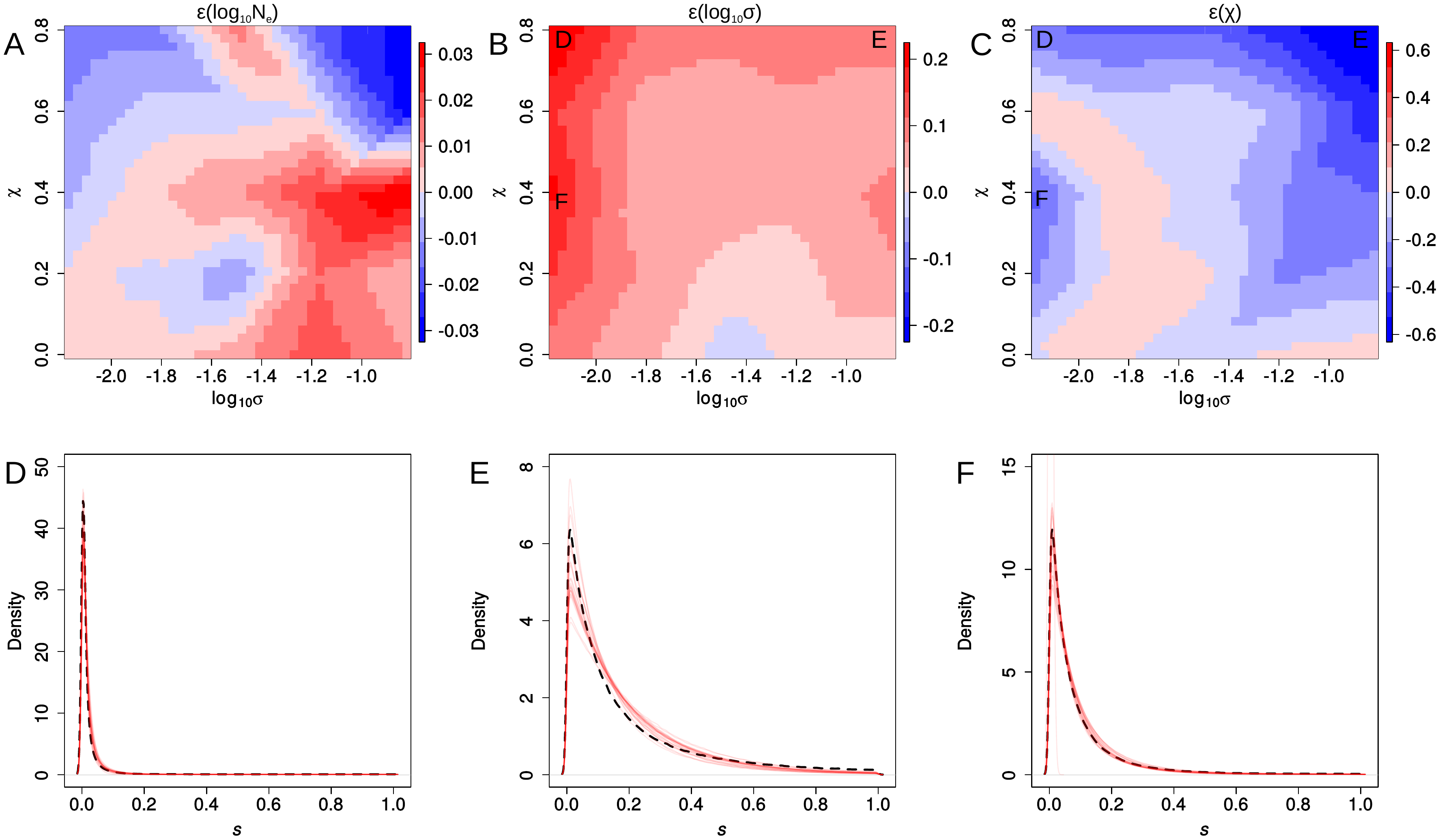}}
\caption{\label{fig_2Derror}\textbf{Accuracy in estimating $N_e$ and DFE parameters $\sigma$ and $\chi$ jointly.} (A,B,C)  Sets of simulations of 100 loci were conducted for combinations of parameters $\sigma$ and $\chi$ over a grid from their prior range and we evaluated the median approximation error ($\epsilon$=estimate-true) over 25 replicates. Color gradients indicate the extent of overestimation (red) or underestimation (blue) of each parameter. These results suggest very high accuracy when estimating $N_e$ with maximum $\epsilon \approx  0.03$ or 1\% of the prior range and rather low for $\sigma$ (about 10\% of the prior range).  In contrast, $\epsilon$ is rather large for $\chi$, spanning up to 75\% of the prior range. This is due to several combinations of $\chi$ and $\sigma$ leading to very similar shapes of the truncated GPD. This is illustrated in panels D, E anf F, where we show the true (dashed black line) versus estimated (red) DFE obtained for 25 replicates using parameter combinations of $\
chi$ and $\sigma$ as indicated in panels B and C.}
\end{figure}

\newpage
\begin{figure}[!htb]
\centering
\centerline{\includegraphics[keepaspectratio,scale=0.65]{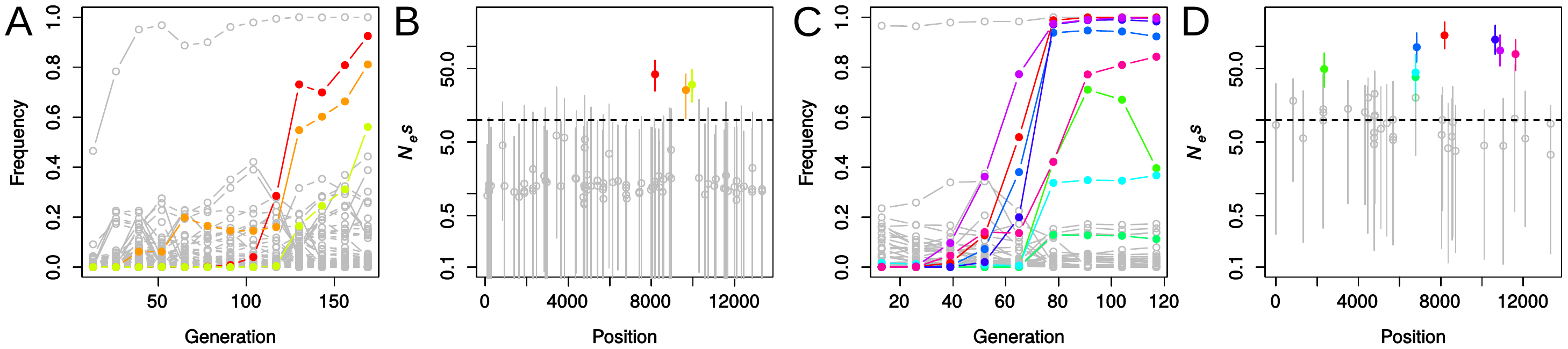}}
\caption{\label{fig_trajectories} \textbf{Allele trajectories and posterior estimates for $N_es$ for control and drug-treated Influenza.} Non-significant loci are colored grey and significant loci are colored with a unique color for each locus.
}
\end{figure}
\end{document}